# Lethal Mutagenesis in Viruses and Bacteria

Peiqiu Chen[1,2], Eugene I. Shakhnovich[1]

[1]Department of Chemistry and Chemical Biology,  [2]Department of Physics,

Harvard University, 12 Oxford Street, Cambridge, MA 02138

**Abstract**

Many (but not all) studies addressed the effect of mutations on populations in the framework of traditional population genetics approaches where beneficial or deleterious mutations were postulated a'prori. Here we study how mutations which change physical properties of cell proteins (stability) impact population survival and growth. In our model the genotype is presented as a set of N numbers – folding free energies of cell's N proteins. Mutations occur upon replications so that stabilities of some proteins in daughter cells differ from those in parent cell by random amounts drawn from experimental distribution of mutational effects on protein stability. The genotype-phenotype relationship posits that unstable proteins confer lethal phenotype to a cell and in addition the cell's fitness (duplication rate) is proportional to the concentration of its *folded* proteins. Simulations reveal that lethal mutagenesis occurs at mutation rates close to 7 mutations per genome per replications for RNA viruses and about half of that for DNA based organisms, in accord with earlier predictions from analytical theory and experiment. This number appears somewhat dependent on the number of genes in the organisms and natural death rate. Further, our model reproduces the distribution of stabilities of natural proteins in excellent agreement with experiment. Our model predicts that species with high mutation rates, tend to have less stable proteins compared to species with low mutation rate.

## Introduction

Mutation rates play an important role in the evolution and adaptation of bacteria and viruses. There is considerable experimental evidence suggesting that high mutation rates in RNA virus populations has powered their rapid evolution (1-4). However, artificially elevated mutation rates were shown to have deleterious effects on the fitness of RNA viruses, and eventually lead to extinction of the viral population beyond certain mutation rate thresholds (5-14). This observation is called lethal mutagenesis for RNA viruses. Several authors proposed to use lethal mutagenesis to cure or control infection with RNA viruses, using certain mutagens (10, 11). The possibility of lethal mutagenesis in bacteria was also suggested and studied recently (15-17).

Previously, many attempts have been made, using population genetics, to theoretically describe the effect of mutation rates on the survival of a population (17-19). The effect has frequently been described within the paradigm of Muller's ratchet, (18-20) where the genome of an asexual organism accumulates stochastic deleterious mutations in an irreversible manner, leading to the systematic decrease in the fitness of the organism. Muller's ratchet applies to finite, asexual populations. It states that if back mutations cannot occur, eventually any finite asexual population will accumulate deleterious mutations and the mutation-free wild type would be lost. While this model has provided some useful insight into the phenomenon of lethal mutagenesis, they often assume a single fitness peak, absence of back or compensating mutations and depend heavily on arbitrary parameters, such as selection coefficients or the deleterious mutation rates. Such analyses therefore lack a more fundamental connection between the physical properties of the proteins within the organism, the metabolic network of the organism, and the feedback relationship between the mutation rate and organismal fitness.

In recent years, several theories of lethal mutagenesis have been proposed (5, 14, 15, 21). In a marked departure from earlier phenomenological approaches Zeldovich et al. (5) suggested a model assuming that the loss of protein stability would lead to loss of essential functions within the organism and therefore to lethal phenotype. The evolution of a population in this model was mapped to a diffusion process in a multidimensional hypercube where each dimension represented stability of an essential gene and adsorbing boundary conditions at $\Delta G = 0$ boundaries (where $\Delta G$ is the difference between free energy of the folded and unfolded proteins which is the thermodynamic measure of protein stability (5)) were imposed to account for the fact that loss of stability confers lethal phenotype on an organism. This model differs from previous approaches in that, instead of depending on arbitrarily calibrated parameters such as selection coefficients or deleterious mutation rates, it is based only on the statistical distribution of proteins' folding free energy change after point mutations, which had been directly derived from in vitro experiments. Furthermore, it predicts a lethal mutagenesis threshold that is consistent with experimental findings and provides valuable insight into the connections between the fundamental biophysical properties of proteins, the mutation rate, and organismal fitness.

However, despite these insights, the model proposed by Zeldovich et al. (5) is based on a number of simplifying assumptions. First, it assumes a uniform mutation supply within the population; meaning that at any time, mutations could occur in any organism in the population with an equal and constant probability. However, in real biological systems mutations are coupled to duplication. That is, mutations should only occur in organisms during the organismal duplication process. While the formalism developed by Zeldovich et al (5) allows to address the case when mutations are coupled to duplication (see Methods in (5)) this formalism gives numerically accurate predictions for the coupled mutation-duplication cases only in the limit of low mutation rates. However lethal mutagenesis occurs when mutation rate is relatively high (approximately six mutations per genome per replication according to (5, 22)). Second, the model developed in (5) assumes a very simple ''$\Theta$-function-like'' fitness landscape whereby fitness is the same for all protein stabilities as long as proteins are stable, i.e. it is ''flat'' for all $\Delta G < 0$. However in reality as proteins become less stable they spend greater fraction of time in unfolded state reducing therefore the effective concentration of functional proteins, which affects fitness. Our study overcomes these limitations in a new computational model as outlined below.

If the organism has conservative duplication mechanism, as is the case for RNA viruses, then mutations would only occur, with certain probabilities, in the descendent copy, while the parent copy would remain unchanged (that may also be the case in organisms with double-stranded genomes where methylation mechanism keeps master copy of the genome preserved). If the organism has semi-conservative duplication, as in bacteria and DNA viruses and double-stranded RNA viruses, then mutations could happen with certain probabilities to both the parent copy and the descendent copy during the replication process.

Here we present a detailed study of effect of mutations changing protein stability on the fitness, based on the coupled mutation-duplication scenario and more explicit consideration of the effect of protein stability on fitness. Simulating evolution and population growth in this model we observe lethal mutagenesis in conservative and semi-conservative replicating mechanisms. Further, we show how stationary distribution of protein stabilities (folding free energies) emerges and discuss the physical and evolutionary reasons for the observed moderate stability of proteins.

**The Model**

In our model, we assume that the duplication rate of an organism depends on the functionality of each of the proteins involved in the organismal duplication process. The number of duplication controlling proteins does not necessarily involve all the proteins in the organism. Instead, it could be a small or large subset of the total number of proteins in the organism, and the number may vary between species and strains (23). However, the organism is only able to replicate efficiently when all of these proteins are able to function properly (24). Failure of a single duplication-controlling protein might result in the organism's dysfunction and hence reduce the organismal duplication rate.

Since a lack of stability in each duplication-controlling protein is detrimental to the duplication process, we assume the organism's replication rate is directly proportional to the percentage of time spent by each of the duplication-controlling proteins in their respective folded conformations. Protein domains are known to fold thermodynamically two-state (25-27) which means that only unfolded and folded forms of a protein domain can be (meta)stable and that the thermodynamic probability to be in the native state is:

$$f_i = \frac{e^{-\frac{G_f^i}{kT}}}{e^{-\frac{G_f^i}{kT}} + e^{-\frac{G_u^i}{kT}}} = \frac{1}{1+e^{-\frac{\Delta G_i}{kT}}} \qquad (1)$$

where $G_f^i$ and $G_u^i$ are free energies of the folded and unfolded forms of protein i respectively and $\Delta G_i = G_f^i - G_u^i$ is the difference between the two representing the stability of protein i. Factors $f_i$ determine effective concentrations of *functional* duplication controlling proteins in the organism. For simplicity, here we assume that the N proteins regulating organism duplication are independent of each other; therefore the percentage of folded proteins should be a function of only their stabilities $\Delta G_{fu}^i$ and the environmental temperature.

The organismal fitness–growth rate in a fixed environment is therefore determined by the percentage of properly folded duplication controlling proteins in the organism. The duplication rate for an organism with N different proteins controlling the duplication process can be expressed as:

$$b = b_0 \prod_{i=1}^{N} f_i \qquad (2)$$

here, $b_0$, is the overall birth rate, a constant parameter that is determined by other organismal environmental factors such as temperature, acidity, or nutrition and by expression levels of all duplication controlling proteins (which are assumed constant in this model).

It is clear from equation (1) that greater absolute value of (negative) protein folding free energy provides a higher ratio of the folded proteins to the unfolded ones in the organism and therefore a higher effective concentration of folded, functional protein. At given environmental conditions, the folding free energies of proteins depend on their sequences, i.e., the genotype of an organism. As these quantities are of principal importance for our study, the genotype of an organism enters our analysis through the set of folding free energies for the duplication-controlling proteins, denoted as $\{\Delta \vec{G}\} = (\Delta G_1, \Delta G_2, \Delta G_3 ... \Delta G_N)$

The effect of a mutation on a protein's stability can be viewed as a random change. By introducing a new mutation, the folding free energy of the original protein is altered. The magnitude of the change is treated as stochastic in our model, whereas experimental evidence (5, 28, 29) has shown that the impact of random mutations on protein stability is biased towards the deleterious side; it has a Gaussian-like distribution with an average destabilizing effect of around 1kcal/mol, and standard deviation of around 1.7kcal/mol (5, 28).

The value of fixed death rate can be highly dependent on the species under study. Although the possibility of bacteria aging has been explored, their natural death rate is small, and the aging process would take a long time (30, 31). However, for viral particles, because of various lytic and non-lytic immune systems (32), their death rate can vary from a small fraction to the majority of the viral population. Therefore, in the following discussion, we will consider a small or absent natural death rate for bacteria and focus on the death rate influence on viral particles.

In addition to the duplication and mutation factors, we also recognize that from various essential gene experiments, the failure for any key protein to fold in a cell can be lethal to the organism. Therefore, we assume, as in (5), that loss of stability for any protein (in which case the folding free energy for this protein is greater than 0) causes the death of an organism (i.e., confers a lethal phenotype). In our analysis this biological requirement is cast as a boundary condition (5):

$$P(\Delta G_1 ... \Delta G_i, ... \Delta G_N, t) = 0 \text{ if } G_i \geq 0 \quad i = 1 ...... N \quad (3)$$

Here $P$ is the number of organisms in the population which have genotype $\{\Delta \vec{G}\} = (\Delta G_1 ... \Delta G_i, ... \Delta G_N)$ at time t.

The relation (3) indicates that there are no organisms in the population whose genomes encode unstable proteins (we focus here on viruses and bacteria which supposedly do not contain intrinsically disordered proteins (33)). The biological motivation for this condition is that not only expression of unstable proteins deprives the organism of functional proteins but also may lead to aggregation of expressed unfolded proteins causing the death of an organism.

### Simulation of evolution of populations.

Based on these principles, we simulate the evolution of populations for semi-conservative and conservative duplications. First, we prepare initial species with 100 identical organisms of the same genotype; stabilities of N proteins in each organism constituting the initial population have random values drawn from the analytical distribution of the functional form described in our previous work (5). At each time step an organism can duplicate with probability determined by the genotype-dependent duplication rate parameter given by Eq.2. An organism is eliminated as soon as mutation occurs which confers any of its proteins a folding free energy value greater than zero.

Upon duplication, mutations may happen in a descendant organism. Mutation in our model represents the change in stability of one or more proteins in the daughter organism compared with parent organism, i.e. genotype of the daughter organism can be presented as

$$\{\Delta\vec{G}\}_{daughter} = \{\Delta\vec{G}\}_{parent} + \{\Delta\Delta\vec{G}\}$$

where $\{\Delta\Delta\vec{G}\} = (\Delta\Delta G_{i_1}, \Delta\Delta G_{i_2}....\Delta\Delta G_{i_s})$ describes changes of stabilities upon a duplication event which resulted in $s$ mutations in proteins $(i_1, i_2....i_s)$. For semi-conservative duplication, mutations might occur in both the parent copy and the descendent copy. If it is conservative duplication, mutations would then occur in the descendent copy only. We generate the number of mutations $s$ at each duplication effect in a daughter organism, according to a Poisson distribution, and the parameter of the Poisson distribution $m_{organism}$ is the average number of mutations per genome per duplication, for this particular species. The mutation rate for each gene in each copy is then $m_{gene} = m_{organism}/N$. After selecting $s$ - the total number of proteins to be mutated at a given replication event - we decide which proteins to be mutated by selecting the set $(i_1, i_2...i_s)$ at random. When a mutation occurs in a protein, the protein's sequence would be physically changed. In this study we do not consider protein sequences explicitly. Rather we posit that free energy of the mutant protein might have a different folding free energy, and the free energy difference $\Delta\Delta G_i$ between wild-type and mutant protein is a random value drawn from a distribution based on statistics of free energy changes collected in multiple protein engineering experiments. To this end, we determine the statistics of changes of protein stability upon mutations from the ProTherm database (28). This database contains information on more than three thousand point mutations, across all currently performed point mutation experiments. The statistics show that protein folding stability change due to point mutation roughly forms a Gaussian distribution, where the mean is 1kcal/mol and the standard deviation is 1.7kcal/mol. Therefore, when mutation occurs, we alter the protein stability by an amount drawn from this Gaussian distribution. The mutant daughter organism will therefore have an altered fitness value (derived from (Eq. (2)), due to the altered stability of some of its proteins.

For viruses with a certain death rate, we set the death rate as a constant with respect to time. At each generation, depending on the death rate, we randomly kill a certain fraction of new organisms. We also impose an upper limit of population size of 5000 organisms by culling excess organisms at random. Thus population dynamics in our simulations essentially maps into a peculiar N-dimensional diffusion process with branching and growth with ''position-dependent'' (Eq.2) branching rate and with proper absorbing boundary conditions as explained above (Eq.3). We have run many series of independent simulations to eliminate the effect of genetic drift caused by this upper limit on the total population. During the numerical simulation, we let organisms evolve in a stable environment, with a time scale of around 5000 generations, and we study the population dynamics and evolution of protein stabilities at various parameters.

**Results**
**Lethal Mutagenesis for Bacteria and DNA Viruses**

With these physiological assumptions, it is apparent that for DNA based organisms with semi-conservative duplication, a mutation rate upper threshold must exist for a sustainable population growth. This is because mutation can happen to both copies and, when duplication happens, both the parent and the descendent copy could undergo a fitness increase or decline, depending on the change of certain protein stabilities. We hereby define our lethal mutagenesis threshold as the minimum mutation rate that can result in a species with sufficient initial large population to eventually go extinct. Our simulated results showed that for an organism with a sufficiently high mutation rate, the species is not stable and may eventually become extinct. However, as shown in Figure 1 as well as in previous work (5), as long as the mutation rate is smaller than some critical value, the population can have a sustainable growth for a sufficiently long time.

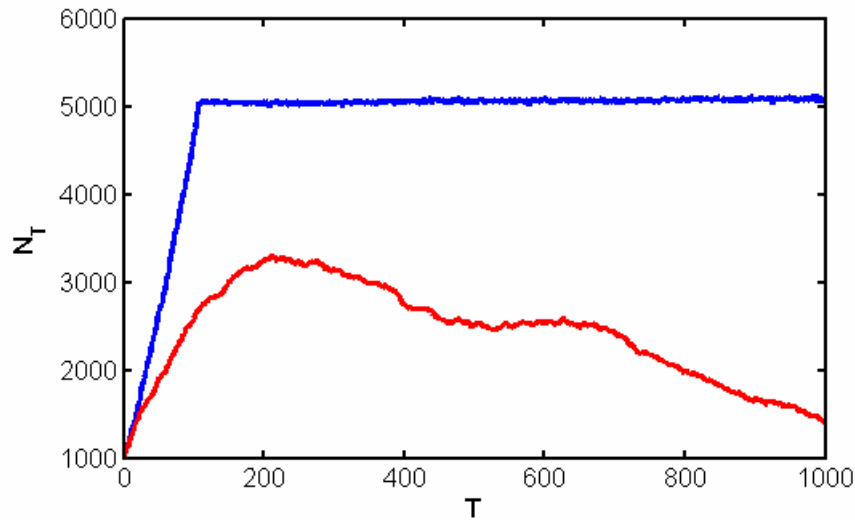

**Figure 1**. *Population dynamics with different mutation rates for a bacterium with 20 duplication-controlling proteins (Red line, 4 Mutations/Duplication; Blue Line, 3 Mutations/Duplication). It is apparent that when the mutation rate is high, the population becomes extinct very quickly, whereas when the mutation rate is low, the population can have positive growth.*

The effect of mutations on population dynamics depends on the number of duplication controlling proteins N as shown in Figure 2.

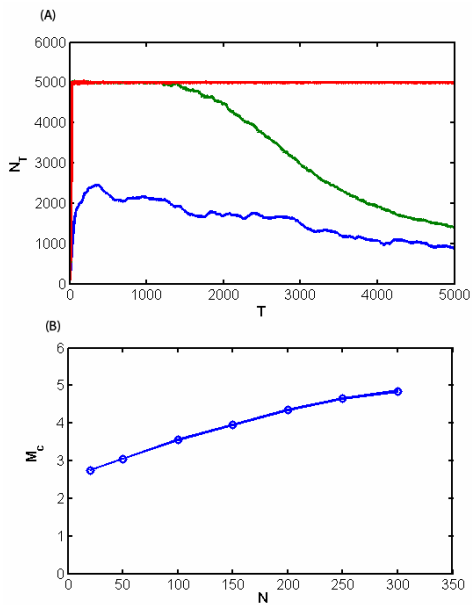

***Figure 2.*** *Population dynamics depends on the number of duplication-controlling proteins. **A**. Population dynamics at fixed mutation rate of 3*

*mutations per organism per duplication. (Blue line: 10 proteins; green line, 50 proteins; red line, 100 proteins). When the duplication-controlling protein number is low, the population becomes extinct very quickly, whereas when the protein number is high, the population can have sustainable growth.* **B** *Mutagenesis threshold $m_c$ as a function of number duplication-controlling protein number for semi-conservative duplication. As the number of duplication-controlling proteins increases, the mutagenesis threshold grows as well. The threshold shows tendency of saturation with increased number of duplication-controlling proteins.*

The lethal mutagenesis threshold shown in Fig.3B is slightly different numerically from that predicted by the previous analytical model (around 3 mutations per genome per replication for semi-conservative duplication and around 6 for conservative one for all genome sizes)(5), and the reasons may be two-fold. Firstly, herein we consider the coupled duplication-mutation scenario, in which mutations only occur during duplication events. Secondly, we also consider the more realistic duplication rate scenario, in which the stabilities of proteins affect the organism duplication rate; that is, all duplication controlling proteins have certain influence on the rate of duplication. The more properly folded copies of a protein are there in the organism, the faster the duplication rate and hence higher the organism fitness. Hence there is a collective contribution term from all the duplication-controlling proteins in the organism. As shown in Figure 2A, at fixed mutation rate of 3 mutations per genome per duplication, increase in the number duplication controlling proteins results in a lower probability that the species will become extinct. This can also be inferred from the fitness landscape in our model (Eq.2). If there are more proteins controlling the duplication process, a slight change in one of them would likely not have such a dramatic effect on the overall duplication rate. In contrast, if the duplication rate is determined by only a few proteins, dysfunction or lack of stability in any one of them would lead to either a great decrease in the duplication rate or the death of the organism. This feature of the model is different from our early study which considered flat fitness landscape for stable proteins (5)

A control simulation has been performed using the fitness landscape exactly as in (5) where duplication rate is a constant as long as all proteins are at least marginally stable. In this constant duplication rate case, rather than observing the lethal mutagenesis threshold gradually increasing with the increase of duplication-controlling protein number, the lethal mutagenesis threshold is a constant value of around 2.5, essentially independent on the number of duplication-controlling proteins (Figure 3). This finding is consistent with (5) which also predicted that lethal mutagenesis threshold (counted as number of mutations per genome per generation) is independent on the genome size. The reason for slight numeric discrepancy of 20% between analytical predictions and simulations is due to the effect of coupling of mutations and duplications.

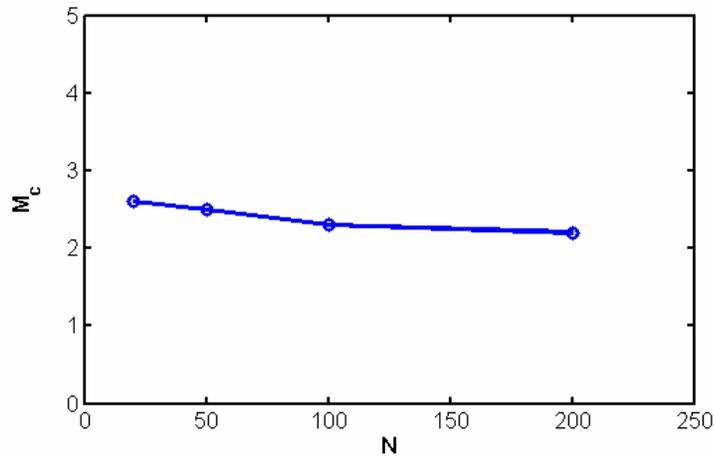

*Figure 3.* *Mutagenesis threshold $m_c$ as a function of the number of proteins controlling the duplication process, in the control simulation whereby the duplication rate is set to constant independent on protein stabilities. Here the lethal mutagenesis threshold is roughly constant, and about 20% lower than the analytical prediction because of the coupled mutation-duplication effect in this simulation.*

Therefore, our results, as indicated in Figures 2B and 3 suggest that even under optimal growth conditions for bacteria,, when the natural death rate is negligible, there still exists a certain lethal mutagenesis threshold for the species. Moreover, our results in Figure 2B indicate that the lethal mutagenesis threshold increases and plateaus when the number of proteins controlling the duplication process increases.

In addition to the semi-conservative duplication mechanisms described above, we recognize that DNA viruses should have a greater death rate in their population due to various lytic and non-lytic immune mechanisms in their host cells (32). Therefore, in order to account for this effect, we incorporate a death rate variable in our model; this death rate can range from 10% of the initial average population birth rate to 90% of the initial average population birth rate. As we can see in Figure 4, in the presence of this finite death rate, the lethal mutagenesis threshold is significantly reduced. As natural death rate increases from 10% to 90% of the duplication rate, the lethal mutagenesis threshold for a population of organisms with 20 duplication-controlling proteins decreases from 2.8 mutations per duplication to around 0.5 mutations per duplication. (see Figure 5).

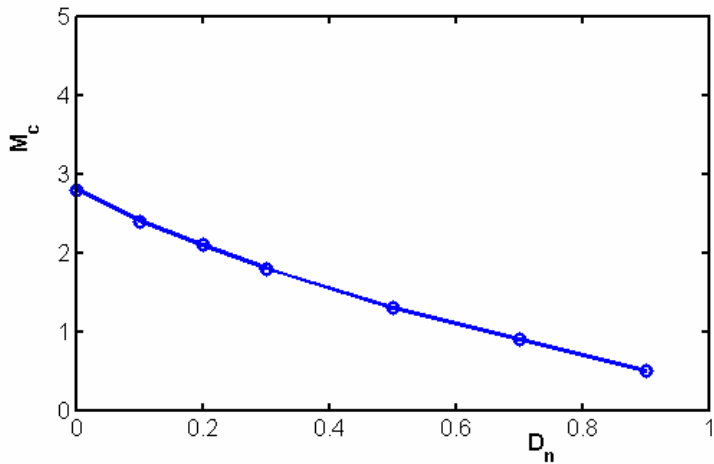

*Figure 4. Lethal mutagenesis threshold $m_c$ as function of $D_n$ - natural death rate for DNA viruses. This simulation is performed for organisms with 20 duplication-controlling proteins. The natural death rate is measured as a ratio probability of death of a progenitor at each duplication event to the birth rate.*

**Lethal Mutagenesis for RNA Viruses**

One significant difference between RNA virus duplication and DNA bacteria and viruses is that most RNA viruses perform conservative duplication instead of semi-conservative duplication. In our numerical simulation, we use a simplified model for RNA viruses in that during genome duplication, one master genome is preserved while the descendent genome might undergo certain mutations. Our results showed that in the case of the preservation of the original stable genome in the RNA virus population during conservative duplications, they can sustain higher mutation rates before reaching lethal mutagenesis threshold compared to that of the DNA based bacteria and viruses. This is also consistent with their evolutionary nature, since due to the lack of error correction mechanisms in the RNA viruses, the descendent copies might undergo certain mutations with a usually higher mutation rate compared to that of the DNA viruses and bacteria.

Meanwhile, like the DNA viruses' survival rate, we also note that RNA viruses' survival depends to a large extent on the various interactions with their host cells; the various lytic and non-lytic immune systems in the cell have a certain impact on the survival rate of the viral population. Since RNA viruses usually have smaller genomes; many of them contain around a dozen of proteins, we therefore consider most of their proteins in the genome as contributing to the organism duplication rate (34). Here we take the number of duplication-controlling proteins as twenty and study the lethal mutagenesis threshold as a function of viral natural death rate. As shown in Figure 5 below, the threshold decreases quickly with the increase of natural death rate. It is also worthwhile to point out that, although previous analytical work showed that in the low mutation rate regime, conservative duplication's mutation rate has the same effect as twice of the semi-conservative duplication's mutation rate (5).

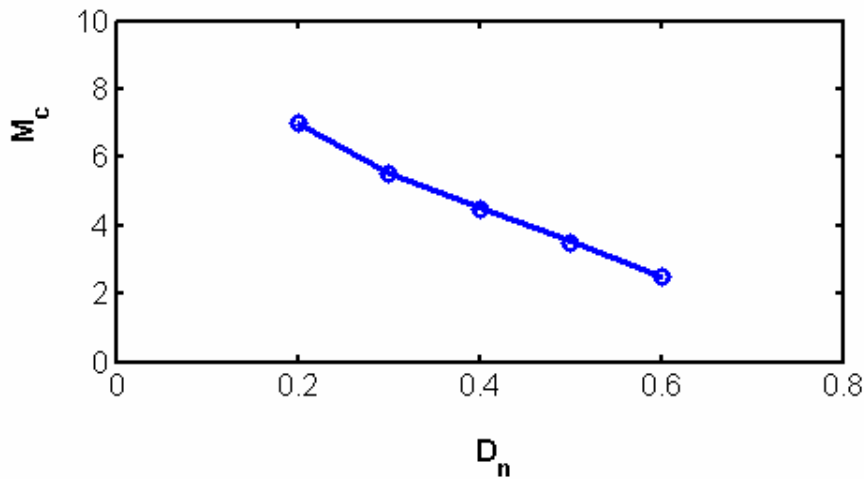

*Figure 5. Lethal mutagenesis threshold as a function of natural death rate for RNA viruses, with duplication-controlling protein number fixed at 20.*

Our numerical simulation actually showed that under the same natural death rate and same numbers of duplication-controlling proteins, the conservative duplication's lethal mutagenesis threshold would be around 15% greater than twice of the semi-conservative duplication mechanism's lethal mutagenesis threshold. Therefore the coupled mutation-duplication scenario is playing a role in conservative replication case by shifting lethal mutagenesis threshold slightly upwards compared to the analytical prediction for the uncoupled case.

**Protein stability distribution in evolved populations.**

After evolving in a steady thermal environment for a sufficient time, the distribution of protein stabilities within an organism will reach its equilibrium state, as shown in Figure 6.

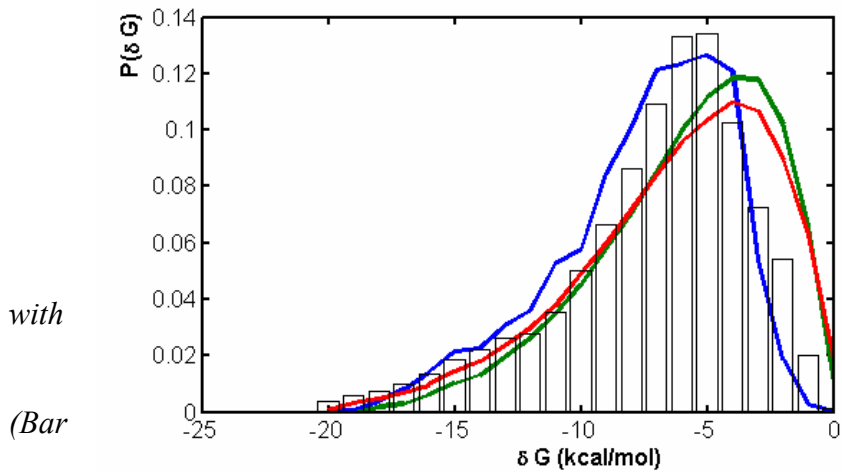

*Figure 6. Distribution of protein stabilities in bacterial species with high and low mutation rates. (Bar Plots: Experimental Data compiled from the ProTherm database(28) ; Blue Line: Low mutation rate case, 0.1 Mut/Dup; Green Line: Analytical model prediction from(5) ; Blue line: High mutation rate case, 2.5 Mut/Dup). The duplication controlling protein number is fixed at 20, and natural death rate is neglected. Compared to species with high mutation rate, low-mutation rate species have significantly more stable proteins. At low mutation rate, the stability distribution agrees very well with experimental data.*

An important new observation from our model is that the mutation rate significantly affects the stationary stability distributions for all proteins in a population. We hereby compare protein stability distribution of species with high mutation rate and species with low mutation rate. The results showed distinctively that when mutation rate is increased, more and more proteins become marginally stable. Moreover, as we can observe from Figure 6, that at low mutation rate, which is the case for most DNA based organisms, our numerical result of protein stability distribution agrees very well with the stability distribution of real proteins drawn from the experimental database, which is a considerable improvement compared with the previous analytical model.

Furthermore, recent analysis suggested that RNA virus proteins, may be indeed different from bacterial ones having lower stability and loose packing of viral proteins (35). Our results, as shown in Figure 7, show a uniform increase of protein average folding free energy value with the increase of mutation rate. It is very obvious that species with high mutation rate have less stable proteins, whereas lower mutation rate gives the species less pressure to adapt, therefore resulting in on average more stable proteins.

*Figure 7. The mean protein stability across all organisms in the species versus mutation rate. For 20 duplication controlling proteins. Horizontal axis m denotes the mutation rate, per genome per duplication.*

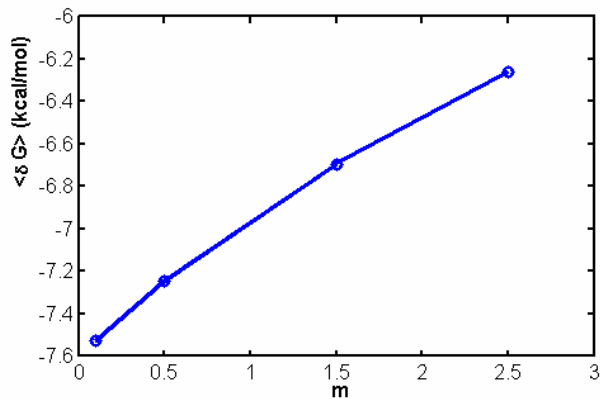

### Discussion

In this work, we systematically studied lethal mutagenesis for several biological systems and discussed the effect of various parameters, such as mutation rate, death rate, duplication-controlling protein numbers, and duplication mechanisms, on the lethal mutagenesis threshold.

Our model is based on the minimal biologically reasonable assumption that the organismal duplication rate is determined by the concentration of functional (i.e. correctly folded) duplication-controlling proteins, and that loss of stability by any of these would result in a lethal phenotype. When all of the duplication-controlling proteins in the organism are stable and properly folded, the organismal duplication rate is high, and the population growth is fast. However, deleterious or beneficial mutations may occur during the duplication process. A deleterious mutation is one that increases the folding free energy of a duplication-controlling protein, destabilizing it and decreasing the number of its properly folded copies in the organism. Hence given the steady expression levels, fewer properly functioning ones remain in the organism after a deleterious mutation. This will decrease the duplication rate and lead the population into a static, non-expanding phase. Furthermore, a deleterious mutation could also result in a protein having folding free energy greater than zero, which is lethal to the organism. In this case, the organism would not be able to survive and would eventually die.

Our result showed that the lethal mutagenesis threshold increases and plateaus with the number of duplication-controlling proteins. On the other hand, the threshold could be reduced with an increase in the natural death rate for the organism. Therefore, our work might potentially lead to more directed search for bacterial lethal mutagenesis (15). Although previous lethal mutagenesis experiments have been more focused on RNA viruses, recently several experiments have also been carried out on bacteria systems with increased mutation rates (36). However, these bacteria usually exhibit unchanged fitness or some beneficial mutation traits, rather than decreased fitness, in the stable environment of the experimental set-up. This is because, although the elevations in the mutation rate in these systems are between 10-fold and 100-fold over their original mutation rates (37), bacteria's natural mutation rate is rather small, and is estimated to be around 0.003 per organism per duplication (38). (The increase of fitness at higher mutation rates which are still well below the lethal mutagenesis threshold was found in recent ab initio microscopic simulations of model cells (39)) Therefore, even, for mutator phenotypes the mutation rates of bacteria are still well below the lethal mutagenesis threshold value and therefore could lead to fixation of beneficial mutations by hitchhiking instead of extinction.

However, lethal mutagenesis is still potentially achievable in bacteria. In our model, we observed that the lethal mutagenesis threshold could be reduced by choosing a bacterium species with fewer duplication-controlling protein numbers, in an environment with a non-negligible natural death rate. For example, for a bacterium or DNA virus species with 20 duplication-controlling proteins and a death rate equal to 20% of the initial duplication rate, the lethal mutagenesis threshold is roughly 2.1 mutations per genome per duplication. Even though this number is still above the current level of the observable bacterial mutation rate, exploring bacteria's lethal mutagenesis in various experimental settings can still lead to basic insights about mutation, replication rate, and organismal fitness; therefore may be well worth the effort.

We also separately discussed lethal mutagenesis in DNA-based organisms and RNA-based organisms. Our results showed that given the same natural death rate, organisms with conservative duplication mechanisms, such as RNA viruses, could tolerate a higher mutation load than DNA-based organisms with a semi-conservative duplication mechanism, in accord with the prediction of the analytical theory (5). Experimentally, ribavirin is the drug commonly considered a candidate for causing lethal mutagenesis in RNA viruses (11, 40). Many experiments have focused on introducing ribavirin to retroviruses such as HIV, hepatitis C, the vesicular stomatitis virus (VSV), etc. This often leads to an increased mutation rate and eventually the lethal mutagenesis of such viruses. As shown by Crotty et al. in 2001 (22), a population of polioviruses can become 90% extinct at exactly six mutations per genome per replication, while the normal mutation rate for this virus is approximately one mutation per genome per replication. This is in agreement with our prediction of around 6.2 mutations per genome per duplication for RNA viruses with 15 proteins controlling the duplication rate and a natural death rate equal to 25% of the wild type's duplication rate.

One of the advantages of our model, compared to that of Muller's Ratchet theory (18), is that instead of assuming various arbitrary parameters for the mutation's effects on fitness or selection coefficients, our model only relies on the fundamental connections between protein stabilities and the organism duplication rate. In recent years, much work has been done based on the original model of Muller's ratchet theory (17, 19, 41). Although these models provided many useful insights on various aspects of adaptation and mutation, our model is different in that we link the fundamental genotypic property – protein stability - with the phenotype and do not employ the selection coefficient and other a'priori unknown arbitrary parameters. Here we infer fitness directly from the stabilities of duplication-controlling proteins of the organism and relate the death phenotype to the loss of function of key proteins.

Recently Bull et al. proposed a model of lethal mutagenesis of bacteria and viruses (14, 15). In their model, the extinction threshold for bacteria is 0.69 deleterious mutations per genome per duplication, and the threshold for viruses is different from ours too. This threshold value differs from the one reported here in that they only considered deleterious mutations, whereas we considered all mutations affecting protein stability, which may be beneficial deleterious or neutral. . In our model, a beneficial mutation makes a duplication-controlling protein more stable, with more copies of properly functioning proteins in the cell. Beneficial mutations constitute a significant fraction among all point mutations in our model, therefore it is very important to take them into consideration. The beneficially mutated strains will then have a higher duplication rate and be farther from the ''death boundary'' and therefore produce more offspring. Meanwhile, we also allow mutations occurring in different proteins to interfere with each other in our model through their cumulative effect on fitness Eq.2, and the dynamics of multiple mutations are taken into account. In contrast to Bull et al.'s earlier works (14, 15), which focus mainly on the abstract parameter of fitness, our model here is based on the thermodynamics and loss of function for the duplication-controlling proteins in the organism; we therefore provide a further linkage between genotype and phenotype.

The effect of protein stability on phenotype was considered in our earlier analytical theory (5). To make the model tractable we did not consider the coupled mutation-duplication process there. Although this approximation works well at low mutation rates, it becomes less quantitatively accurate at high mutation rates. While the analytical theory in (5) provided a qualitatively correct estimate for the lethal mutagenesis threshold, our present study shows that these estimates may be 15-20% off more accurate numbers based on simulations of coupled duplication-mutation events and in a more realistic fitness landscape given by Eq.2. Thus a coupled mutation-duplication model is essential for more accurate estimate of lethal mutagenesis threshold, which occurs at high mutation rates per duplication where the analytical approximation of (5) remains qualitatively but not quantitatively correct.

Many earlier experimental works (10, 22) associated the lethal mutagenesis in RNA viruses with ''error catastrophe'' predicted by Eigen and coworkers for the quasispecies model (42). According to this model at high mutation rates the population loses fitness by delocalizing in sequence space (42). However, it was argued in (5) and demonstrated in (39) that error catastrophe in the quasispecies model is a consequence of a highly unrealistic single fitness peak assumption. On the other hand close agreement between lethal mutagenesis threshold predicted here and in and the observed one (40) makes it highly plausible that the physical cause of lethal mutagenesis is the persistent loss of stability of one or more of essential proteins at higher mutation rates.

Another important result of our model is prediction of distribution of stabilities of proteins in evolved populations, *without adjustable parameters* and in almost perfect agreement with experiment (see Figure 6). Analytical theory predicted the distribution of stabilities which was in good but not perfect agreement with reality. In particular the peak of the predicted distribution was shifted towards lower stabilities compared to experimental one, and analytical distribution overall predicted a greater number of marginally stable proteins than observed in experiment. There could be three possible reasons for the discrepancy between analytical results and experiment: 1) bias in the ProTherm database reflecting overall difficulties of experimental study of marginally stable proteins 2) That the assumption of the analytical theory that evolutionary processes had reached steady state may not be fully justified and 3) That the ''$\Theta$-function like'' fitness landscape used in the analytical theory is too oversimplified by not taking into account the gradual loss of fitness as proteins become marginally stable. The present study indicates that 3) is the right reason for the discrepancy between analytical prediction and experiment. When we took into account that low stability proteins essentially lead to lower fitness due to lower fraction of folded (and therefore functional) essential proteins, the agreement between predicted and observed distribution of stabilities very much improved without introducing additional adjustable parameters. Apparently lower fitness at higher proportion of marginally stable proteins gives rise to additional pressure leading to the shift of the distribution towards somewhat more stable proteins.

Our study, as well as previous analytical work provide a realistic explanation to the observation that most proteins are not exceedingly stable – indeed the distribution of protein stabilities peaks around -5kcal/m (see Figure.7). Several authors attributed such moderate stability to the functional requirement of flexibility which requires lower stability of a protein. As an evidence for the coupling between function and stability a circular argument was suggested that observed stabilities of proteins are not too high. However, experiments show that increasing stability by site-directed mutagenesis does not affect activity of a protein. A number of earlier works which seemed to suggest otherwise have been sometimes seriously misinterpreted. In most cases of these earlier studies (see e.g. (43, 44)) *only active site* residues were mutated and indeed loss of activity accompanied by some stabilization of the protein was observed for several mutations. However this result suggests only that carving active site (often enriched with polar and/or charged amino acids) on a surface of an enzyme may indeed destabilize the protein, i.e. that *active site* is nor optimized for stability. However in order to prove that limited stability of a protein is a prerequisite of its function one has to study the functional effect of stabilizing mutations outside active site and show that these also lead to loss of function. These experiments are necessary to separate the indirect effect of mutations on function through stability alteration from a direct effect on function due to mutations in the active site. When protein was mutated outside the active site the increase of stability did not result in a loss of activity (45). Our study provides another evidence for independence of stability and activity of proteins, by accurately predicting the distribution of proteins stabilities without assuming any relation between the two.

Taverna and Goldstein (46) put forward an argument that protein stability is not too high for entropic (in sequence space) reason: That there are much more sequences corresponding to proteins of moderate stability than to more stable ones. A similar observation was made by Bloom et al in their study of neutral evolution of protein stability (47). While this argument is certainly a correct one, it does not provide an answer of why the distribution of proteins stabilities has a well-defined peak around -5kcal/m rather than being shifted towards less stable proteins whose sequences represent an overwhelming majority. Our study answers this question by showing that distribution of protein stabilities is established as a compromise between two opposing factors: A) physical factors which favor less stable proteins for entropic (in sequence space) reasons outlined in (46) and B) population genetics pressure which favors more stable proteins by eliminating from populations the organisms which carry marginally stable proteins. The organisms whose genotypes encode more marginally stable proteins become eliminated from populations for two reasons: 1) their fitness is lower because fraction of folded proteins is diminished and, more importantly, 2) because organisms with less stable proteins are closer to ''death'' boundary i.e. they leave less progeny.

A new insight coming from this study is the dependence of stability of proteins in a population of organisms on mutation rates. Our simulations predict that species with higher mutation rates (such as RNA viruses) should have less stable proteins than their lower mutations rates counterparts. Recent study provides some evidence to that effect – showing that the attributes of proteins stability such as contact density are significantly different in proteins of RNA viruses than that of DNA-based organisms (35). However, as suggestive as it is the work of Tawfik et al (35) is not a proof of lower stability of RNA viral proteins: A much more comprehensive analysis based on throughput measurements of proteins stabilities in several organisms is required to confirm or falsify this prediction.

In conclusion, based on the thermodynamics of proteins in an organism, we proposed a model for lethal mutagenesis of bacteria and viruses. Our model was able to explain the results of RNA virus lethal mutagenesis experiments, including that certain RNA viruses have a threshold of around six mutations per organism per duplication. Our model also provides possible directions for exploring bacteria lethal mutagenesis experimentally, such as performing experiments on bacteria with fewer duplication-controlling proteins or higher natural death rates. Finally, we note that our model of lethal mutagenesis is minimalist in that it did not account for various important biological processes, such as gene recombination, Darwinian selection due to competition for limited resources, or interaction between an organism and external materials. For example, ribavirin has been demonstrated to have at least three activities in vivo or in cell culture, including inhibition of cell duplication, immunomodulatory effects, and incorporation as a mutagenic nucleoside by the viral RNA polymerase (40). Therefore, further efforts and extensions, such as more explicit consideration of protein function, protein-protein interactions, or the structure-function relationship, would also be valuable to the understanding of lethal mutagenesis in bacteria and viruses.

*Acknowledgements.* This work is supported by the NIH. We thank Konstantin Zeldovich for help at the initial stages of the project and many useful discussions.